\documentclass[aps,prl, groupedaddress,amsmath, twocolumn,showpacs]{revtex4}%
\usepackage{graphicx}
\usepackage{amsmath}
\usepackage{amsfonts}
\usepackage{amssymb}

\begin{document}
\title{Synchronization by Nonlinear Frequency Pulling}
\author{M.~C.~Cross}
\affiliation{Department of Physics 114-36, California Institute of Technology, Pasadena,
California 91125.}
\author{A.~Zumdieck}
\affiliation{Max Planck Institute for Physics of Complex Systems, Noethnitzer Strasse 38,
01187 Dresden, Germany.}
\author{Ron Lifshitz}
\affiliation{School of Physics \& Astronomy, Raymond and Beverly Sackler Faculty of Exact
Sciences, Tel Aviv University, Tel Aviv 69978, Israel.}
\author{J.~L.~Rogers}
\affiliation{HRL Laboratories, LLC, 3011 Malibu Canyon Road, Malibu, California 90265.}
\date{June 28, 2004}

\begin{abstract}
We analyze a model for the synchronization of nonlinear oscillators due to
reactive coupling and nonlinear frequency pulling motivated by the physics of
arrays of nanoscale oscillators. We study the model for the mean field case of
all-to-all coupling, deriving results for the onset of synchronization as the
coupling or nonlinearity increase, and the fully locked state when
all the oscillators evolve with the same frequency.

\end{abstract}

\pacs{85.85.+j, 05.45.-a, 05.45.Xt, 62.25.+g} \maketitle

In the last decade we have witnessed exciting technological advances in the
fabrication of nanoelectromechanical systems (NEMS). Such systems are being
developed for a host of nanotechnological applications, as well as for basic
research in the mesoscopic physics of phonons and the general study of the
behavior of mechanical degrees of freedom at the interface between the quantum
and the classical worlds~\cite{R01,C03}. Among the outstanding features of
nanomechanical resonating elements is the fact that at these dimensions their
normal frequencies are extremely high---recently exceeding the 1GHz
mark~\cite{HZMR03}---facilitating the design of ultra-fast mechanical devices.
Since with diminishing size output signals diminish as well, there is a need
to use the coherent response in large arrays of coupled nanomechanical
resonators (like the ones that were recently fabricated~\cite{BR02,SHSICC03})
for signal enhancement and noise reduction. One potential
obstacle for achieving such coherent response is the fundamental problem of
the irreproducibility of NEMS devices. Clearly, as the size of a resonating
beam or cantilever decreases to the point that its width is only that of a few
dozen atoms, any misplaced atomic cluster dramatically can change the normal
frequency or any other property of the resonator. Thus, it is almost
inevitable that an array of nanomechanical resonators will contain a
distribution of normal frequencies. Here we propose to overcome this potential
difficulty by making use of another typical feature of nanomechanical
resonators---their tendency to behave nonlinearly at even modest amplitudes.
We shall demonstrate here that systems of coupled nonlinear nanomechanical
resonators (like the one we studied recently~\cite{LC03}) can self-synchronize
to one common frequency through the dependence of their
frequencies on the amplitude of oscillation.

The synchronization of systems of coupled oscillators that have a distribution
of individual frequencies is important in many disciplines of
science~\cite{PRK01,SYNC}. The coherent oscillations can be used to enhance
the sensitivity of detectors or the power output from sources, as proposed
here. Synchronization is also important in biological phenomena, for
example the collective behavior in populations of animals, such as the
synchronized flashing of fire flies, and the coherent oscillations observed in
the brain.

Although synchronization is often put forward as an example of the importance
of understanding a nonlinear phenomenon, the intuition for the phenomenon, and
indeed the subsequent mathematical discussion, can often be developed in terms of simple linear
ideas. Even the famous example of Huygens' clocks can largely be understood
\cite{BSRW02} in terms of a linear coupling of the two pendulums through the
common support. The larger damping of the symmetric mode (coming from the
larger, dissipative motion of the support) compared with the antisymmetric mode
tends to lead to a synchronized state with the two pendulums oscillating
antiphase. The nonlinearity in the system is present in the individual
motion of each pendulum in the mechanism to sustain the oscillations, and to reach a full
description of the range of possible states must be included in the analysis. However, even
without this drive the oscillators would still become synchronized through the faster
decay of the even mode, albeit in a slowly decaying state. A second important feature of
the model describing the two pendulums, and of many other models used to show
synchronization, is that the essential coupling between the oscillators is
dissipative, whereas in many physical situations the coupling is mainly reactive.

In our example of the coupled array of nanomechanical oscillators we expect to
see predominantly reactive terms coming from the elastic coupling forces
between the oscillators. Furthermore, rather than the mode-dependent
dissipation mechanism described above, we expect that for our nonlinear
nanomechanical oscillators synchronization will arise from the intrinsically
nonlinear effect of the frequency pulling of one oscillator by another. Thus,
in this paper, we propose and analyze a model for synchronization involving
\textit{reactive coupling\/} between the oscillators, which then leads to
synchronization through \textit{nonlinear frequency pulling}---both effects
must be present for synchronization to occur.

Important advances in the understanding of synchronization have come from
studying a simple model \cite{W67} often known as the Kuramoto model
\cite{K75}. In this model, the oscillators are represented as phase variables,
which in the absence of coupling simply advance at a rate that is constant in
time, but with some dispersion of frequencies over the different oscillators.
The coupling is included as infinite-range, or all-to-all coupling, so that
the model is represented by the equations of motion for the $N$ oscillators
(with the dot denoting a time derivative)%
\begin{equation}
\dot{\theta}_{m}=\omega_{m}+\frac{K}{N}\sum_{n=1}^{N}\sin(\theta_{n}%
-\theta_{m}),\qquad m=1\ldots N. \label{Kuramoto}%
\end{equation}
Here the $\omega_{m}$ are the individual oscillator frequencies taken from a
distribution $g(\omega)$, and $K$ is a positive coupling constant.
The synchronization is captured by a nonzero value of a complex order
parameter $\psi$%
\begin{equation}
\psi=R\,e^{i\Theta}=\frac{1}{N}\sum_{m=1}^{N}z_{m}=\frac{1}{N}\sum_{m=1}%
^{N}r_{m}e^{i\theta_{m}}, \label{OrderParameter}%
\end{equation}
with the magnitude $r_{m}=1$ for the Kuramoto model,

The Kuramoto equation shows rich behavior, including, in the large $N$ limit,
a sharp synchronization transition at a value of the coupling constant
$K=K_{c}$ \cite{K75}, which depends on the frequency distribution of the
uncoupled oscillators $g(\omega)$. The transition is from an unsynchronized
state with $\psi=0$ in which the oscillators run at their individual
frequencies, to a synchronized state with $\psi\neq0$ in which a finite
fraction of the oscillators lock to a single frequency, $\Omega=\dot\Theta$,
equal to the mean frequency of the locked oscillators. The transition at
$K_{c}$ has many of the features of a second order phase transition, with
universal power laws and critical slowing down \cite{K75}, and a diverging
response to an applied force \cite{S88}.

Equation~(\ref{Kuramoto}) is an abstraction from the equations describing most
real oscillator systems, leaving out many important physical features. A
natural generalization is to include the magnitude of the oscillations as
dynamical variables \cite{PRK01}, while adding nonlinearities and considering
reactive as well as dissipative coupling. Thus we are led to the model%
\begin{equation}
\dot{z}_{m}=i(\omega_{m}-\alpha|z_{m}|^{2})z_{m}+(1-\left\vert z_{m}%
\right\vert ^{2})z_{m}+\frac{K+i\beta}{N}\sum_{n=1}^{N}(z_{n}-z_{m}).
\label{AmplitudePhase}%
\end{equation}
The behavior including just nonlinear saturation and dissipative coupling
(i.e.\ setting $\alpha=\beta=0$) was analyzed by Matthews et al. \cite{MMS91}.
We will instead study the case of reactive coupling ($\beta\neq0,K=0$) and
allow for nonlinear frequency pulling ($\alpha\neq0$). This model then has two
parameters: $\alpha$ the strength of the imaginary nonlinear term which yields
the frequency pulling, and $\beta$ the reactive coupling strength. In addition
the probability distribution $g(\omega)$ of the $\omega_{m}$ must be specified.
We will study the case of positive $\alpha$ and $\beta$; for a symmetric distribution
$g(\omega)$ the results are the same changing the sign of both $\alpha$ and $\beta$.

The main focus of this paper is analyzing the behavior of
(\ref{AmplitudePhase}), but first we want to show how such an equation might
arise from the equations of motion of realistic nonlinear coupled
nanomechanical resonators. A possible set of equations describing such a
system of $N$ coupled resonators (similar to the system we studied recently in
a different context~\cite{LC03}) is%
\begin{multline}
\ddot{x}_{m}+(1+\delta_{m})x_{m}-\nu(1-x_{m}^{2})\dot{x}_{m}-ax_{m}%
^{3}\label{Eq_DV}\\
-D[x_{m}-\tfrac{1}{2}(x_{m+1}+x_{m-1})]=0.
\end{multline}
The first two terms describe uncoupled harmonic oscillators, where the
coordinate $x_{m}$ measures the position of the $m^{th}$ nanomechanical
cantilever or beam, oscillating in its fundamental mode of vibration. We
suppose the uncoupled oscillators have a linear frequency that is near unity
(by an appropriate scaling of time) so that $\delta_{m}\ll1$.
The third term is a \emph{negative} linear damping, which represents some
unspecified energy source to sustain the oscillations, and positive nonlinear
damping, so that the oscillation amplitude saturates at a finite value. This
saturation value is chosen to be of order unity by an appropriate scaling of
the displacements $x_{m}$. The first three terms comprise a set of uncoupled
\emph{van der Pohl oscillators}. The term $ax_{m}^{3}$ is a reactive nonlinear
term that leads to an amplitude dependent shift of the resonant frequency, observed
experimentally in many nanomechanical resonators~\cite{C00,BR01}. With $\nu=0$
this would give us a set of uncoupled \emph{Duffing oscillators}. The final
term is a nearest neighbor coupling between the oscillators, depending on the
difference of the displacements. This is a reactive term, typical for either
elastic or electrostatic interaction between resonators that conserves the
energy of the system. Others \cite{AEK90} have considered nonlinear
oscillators coupled through their velocities; this is a dissipative coupling
that would lead to $K\neq0$ in the amplitude-phase reduction.

The complex amplitude equation (\ref{AmplitudePhase}) holds if the parameters
$\nu,\alpha,D,\delta_{m}$ are sufficiently small. In this case the equations
of motion are dominated by the terms describing simple harmonic oscillators at
frequency one. We may then write%
\begin{equation}
x_{m}\simeq z_{m}(t)e^{it}+c.c.+\cdots, \label{Eq_Ansatz}%
\end{equation}
where $z_{m}(t)$ is slowly varying compared with the basic oscillation
frequency of unity, and $\cdots$ are correction terms. Substituting
(\ref{Eq_Ansatz}) into the equations of motion (\ref{Eq_DV}) and requiring
that secular terms proportional to $e^{it}$ vanish yields the amplitude
equations
\begin{multline}
2\dot{z}_{m}=(\nu+i\delta_{m})z_{m}-(\nu+3ia)\left\vert z_{m}\right\vert
^{2}z_{m}\label{DuffingAmp}\\
-iD[z_{m}-\tfrac{1}{2}(z_{m+1}+z_{m-1})].
\end{multline}
With a rescaling of time $\bar{t}=\nu t/2$ (\ref{DuffingAmp}) reduces to our
model~(\ref{AmplitudePhase}) except that in our model the nearest neighbor
coupling is replaced by the all-to-all coupling convenient for theoretical analysis.

Since we are interested in the behavior of (\ref{AmplitudePhase}) for a large
number of oscillators, it is convenient to go to a continuum description,
where we label the oscillators by their uncoupled linear frequency
$\omega=\omega_{j}$ rather than the index $j$, $z_{j}\rightarrow z(\omega)$.
Introducing the order parameter (\ref{OrderParameter}), the oscillator
equations can be written in magnitude-phase form as%
\begin{align}
d_{t}\bar{\theta}  &  =\bar{\omega}+\alpha(1-r^{2})+\beta Rr^{-1}\cos
\bar{\theta}\label{theta}\\
d_{t}r  &  =(1-r^{2})r+\beta R\sin\bar{\theta} \label{r}%
\end{align}
where $\bar{\theta}=\theta-\Theta$ is the oscillator phase relative to that of
the order parameter, and $\bar{\omega}$ is the bare oscillator frequency
measured relative to $\Delta,$which is the order parameter frequency
$\Omega=\dot{\Theta}$, shifted by $-(\alpha+\beta)$
\begin{equation}
\bar{\omega}=\omega-\Delta;\ \Delta=\Omega+\alpha+\beta. \label{define_delta}%
\end{equation}
At each time $t$ the oscillators are specified by $\rho(r,\bar{\theta}%
,\bar{\omega},t)$, the distribution of oscillators at shifted frequency
$\bar{\omega}$ over magnitude and phase values. The order parameter is given
by the self-consistency condition%
\begin{equation}
R\,=\left\langle re^{i\bar{\theta}}\right\rangle =\int d\bar{\omega}\bar
{g}(\bar{\omega})\int r\,dr\,d\bar{\theta}\rho(r,\bar{\theta},\bar{\omega
},t)re^{i\bar{\theta}}.\, \label{SelfConsistency}%
\end{equation}
where $\bar{g}(\bar{\omega})$ is the distribution of oscillator frequencies
expressed in terms of the shifted frequency $\bar{\omega}$. Note that unlike
the cases of the Kuramoto model and (\ref{AmplitudePhase}) with $\alpha
=\beta=0$ the imaginary part of this condition is not trivially satisfied even
for the case of a symmetric distribution $g(\omega)$, and in fact serves to
determine the frequency $\Omega$ of the order parameter. Furthermore, this
frequency is not trivially related to the mean frequency of the oscillator distribution.

To uncover more fully the behavior of our model (\ref{AmplitudePhase}) we
consider two issues: the existence of a fully locked state for large values of
$\alpha\beta$; and the onset of synchronization, detected as the linear
instability of the unsynchronized $R=0$ state.

First we look at the fully locked solution for a bounded distribution of
frequencies of width $w$.
We define any state with an $O(1)$ magnitude of the order parameter $R$ as \emph{synchronized}.
If \emph{all} of the phases of a synchronized state are rotating at the order parameter frequency
we call the state \emph{fully locked}.
The solutions are defined by setting
$d_{t}r=0$ which gives%
\begin{equation}
(1-r^{2})r=-\beta R\sin\bar{\theta}, \label{rConstant}%
\end{equation}
and $d_{t}\bar{\theta}=0$, which with (\ref{rConstant}) can be written%
\begin{equation}
\bar{\omega}=F(\bar{\theta})=\beta Rr^{-1}(\alpha\sin\bar{\theta}-\cos
\bar{\theta}), \label{Locked}%
\end{equation}
where the solution to the cubic equation (\ref{rConstant}) for $r$ is to be
used to form the function of phase alone $F(\bar{\theta})$. The function
$F(\bar{\theta})$ acts as the force pinning the locked oscillators to the
order parameter. A particular oscillator, identified by its shifted frequency
$\bar{\omega}$, may be locked to the order parameter if (\ref{Locked}) has a
solution $\bar{\theta}=F^{-1}(\bar{\omega})$ and if this solution is stable.
The stability is tested by linearizing (\ref{theta},\ref{r}) about the
solution. The fully locked solution will only exist if stable, locked
solutions to (\ref{Locked}) exist for \emph{all} the oscillators in the
distribution. In addition, the self consistency condition
(\ref{SelfConsistency}) must be satisfied.

For large values of $\alpha\beta$ the phases of the locked oscillators cover a
narrow range of angles. The imaginary part of the self consistency condition
(\ref{SelfConsistency}) shows that the range of phases must be around
$\bar{\theta}=0$, and (\ref{Locked}) becomes (note $r\simeq1$ here)
\begin{equation}
\bar{\omega}\simeq-\beta R(1-\alpha\bar{\theta}). \label{AlphaLarge}%
\end{equation}
The imaginary part of the self-consistency condition reduces to $\left\langle
\bar{\theta}\right\rangle =0$ (the average is over the distribution of
frequencies), and the real part to simply $R\simeq1$. Finally, averaging
(\ref{AlphaLarge}) over the distribution of frequencies fixes the order
parameter frequency $\Omega\simeq\left\langle \omega\right\rangle -\alpha$.
This construction remains valid for large $\beta$, so that unlike the case
studied by Matthews et al.\ \cite{MMS91}, \textquotedblleft amplitude
death\textquotedblright\ does not necessarily occur at large values of the
coupling constant. The extension of this calculation to find the boundary of
the fully locked state will be presented elsewhere.

We now turn our attention to the initial onset of partial synchronization from
the unsynchronized state. This is calculated by linearizing the distribution
$\rho$ around the unsynchronized distribution which is a uniform phase
distribution at $r=1$, and seeking the parameter values at which deviations
from the uniform phase distribution begin to grow exponentially. Care is needed in
the analysis due to the important role the magnitude perturbations
play.

Introducing the small expansion parameter $\varepsilon$ characterizing the
small deviations from the unsynchronized state, we write%
\begin{equation*}
\rho(r,\theta,\bar{\omega},t)\simeq(2\pi r)^{-1}\delta\lbrack r-1-\varepsilon
r_{1}(\bar{\theta},\bar{\omega},t)][1+\varepsilon f_{1}(\bar{\theta}%
,\bar{\omega},t)],%
\end{equation*}
as well as $R\simeq\varepsilon R_{1}$, with $r_{1},f_{1},R_{1}\propto
e^{\lambda t}$ with $\lambda$ the growth rate of the linear instability. With
this expansion $\rho$ remains normalized to linear order in
$\varepsilon$ providing the average of $f_{1}$ over $\bar{\theta}$ is zero.
The dynamical equations (\ref{theta}-\ref{r}) at $O(\varepsilon)$ lead to the
explicit solutions $r_{1}=R_{1}(A\cos\bar{\theta}+B\sin\bar{\theta})$ with
\begin{equation*}
A=-\beta\frac{\bar{\omega}}{\bar{\omega}^{2}+(\lambda+2)^{2}},\ B=\beta
\frac{(\lambda+2)}{\bar{\omega}^{2}+(\lambda+2)^{2}}, \label{AB}%
\end{equation*}
and $f_{1}=R_{1}(C\cos\bar{\theta}+D\sin\bar{\theta})$, with%
\begin{equation*}
C=\beta\frac{2\alpha(\lambda^{2}+2\lambda-\bar{\omega}^{2})-\bar{\omega}%
[\bar{\omega}^{2}+(\lambda+2)^{2}]}{\left(  \bar{\omega}^{2}+\lambda
^{2}\right)  \left[  \bar{\omega}^{2}+(\lambda+2)^{2}\right]  }, \label{C}%
\end{equation*}%
\begin{equation*}
D=\beta\frac{4\alpha\bar{\omega}(\lambda+1)+\lambda\lbrack\bar{\omega}%
^{2}+(\lambda+2)^{2}]}{\left(  \bar{\omega}^{2}+\lambda^{2}\right)  \left[
\bar{\omega}^{2}+(\lambda+2)^{2}\right]  }. \label{D}%
\end{equation*}
The self-consistency condition (\ref{SelfConsistency}) to first order in
$\varepsilon$ gives%
\begin{equation}
1=\frac{1}{2}\int d\bar{\omega}\,\bar{g}(\bar{\omega})\left[
(A+C)+i(B+D)\right]  , \label{ParPerpInt}%
\end{equation}

\begin{figure*}
[tbh]
\begin{center}
\includegraphics[
height=2.6in,
width=3.0in
]%
{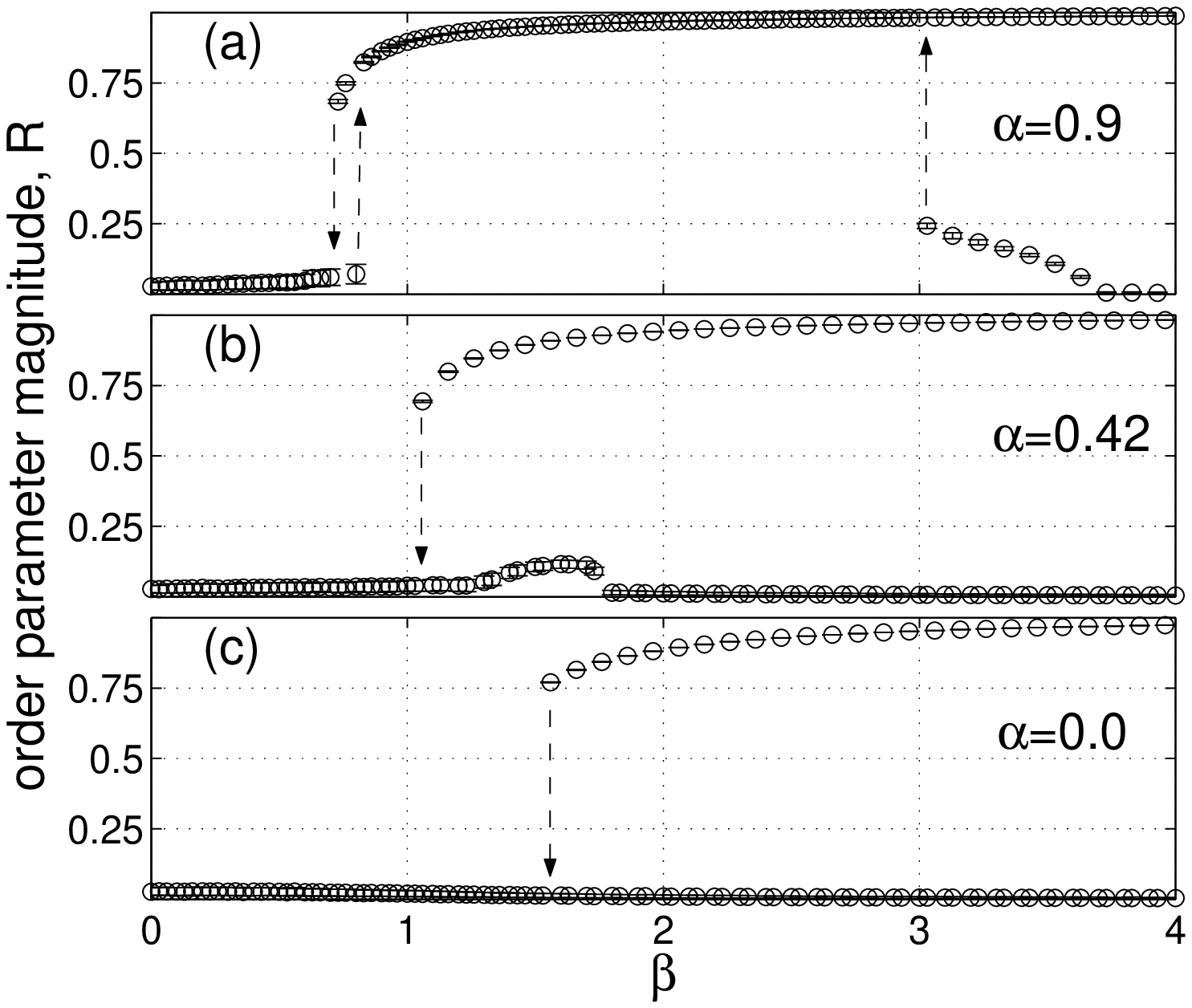}%
~~~~
\includegraphics[
height=2.6in,
width=3.2in
]%
{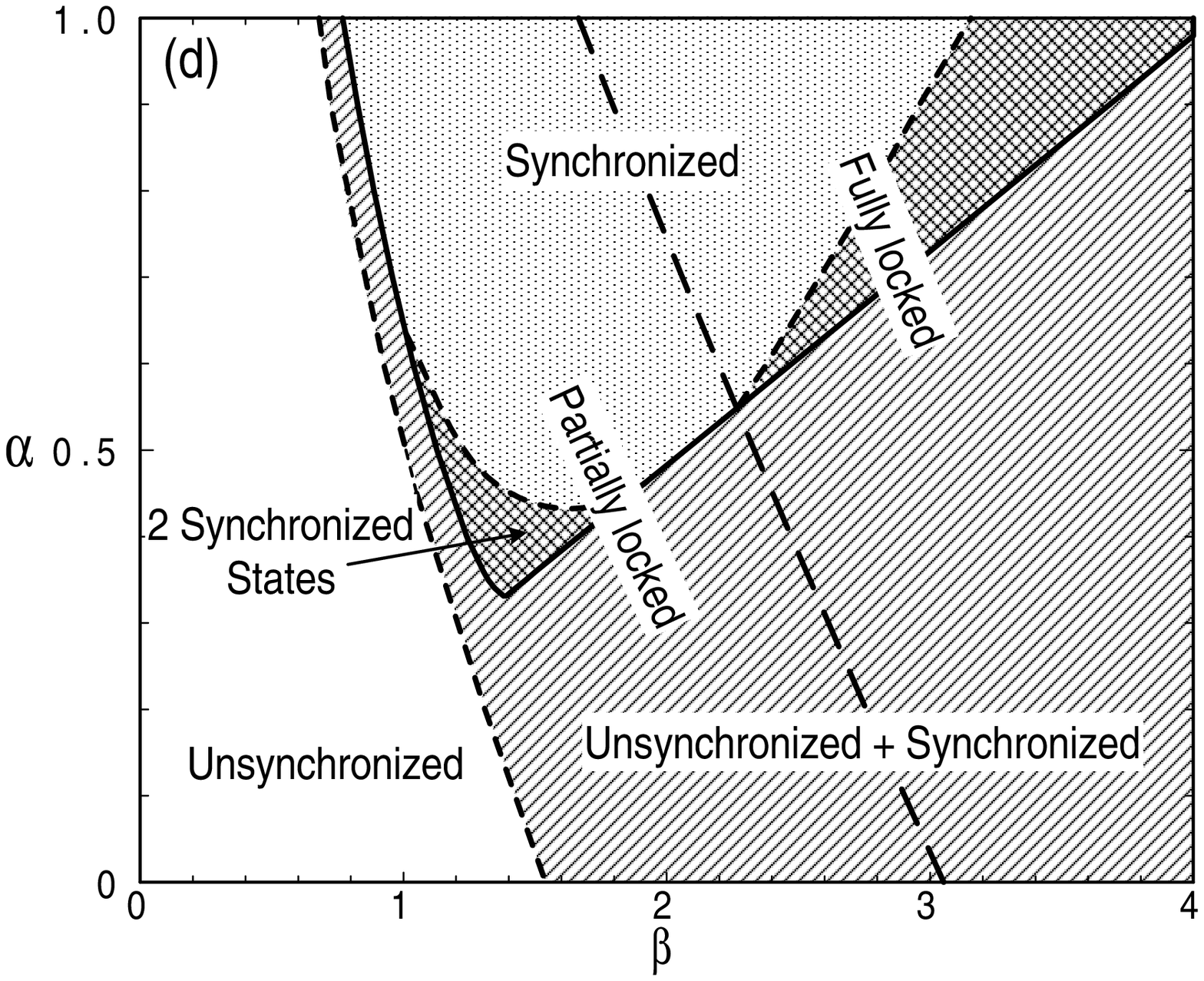}
\caption{Results for a triangular distribution of full width $w=2$. Panels
(a)-(c) show the order parameter $R$ found in numerical simulations of $1000$
oscillators for sweeps increasing and decreasing $\beta$ and for three representative
values of $\alpha$. The symbols are time-averaged values and the error bars
are the standard deviations in $R$ over the averaging time.
Panel (d) shows the phase diagram deduced from sweeps at many
values of $\alpha$, and numerical calculations of the linear instabilities:
solid line - linear instability of the unsynchronized state; short-dashed line -
saddle-node line deduced from jumps of $R$ in the
numerical simulations (denoted by arrows in panels a-c); long dashed line - linear instability
of the fully locked solution (the large $R$ solution is fully locked to the right of this line).}%
\label{Results}%
\end{center}
\end{figure*}
We evaluate (\ref{ParPerpInt}) at the onset of instability where the growth
rate $\lambda\rightarrow0$ (it is not sufficient to put $\lambda=0$ since some
terms of the integrals then diverge). We have evaluated the integrals
analytically for uniform, triangular, and Lorentzian distributions of bare frequencies.
Here we present results for the triangular distribution, for which
the resulting equation for the critical values of
$\alpha,\beta$ and the order parameter frequency at onset must be solved
numerically.

Figure \ref{Results} shows comprehensive results for a triangular distribution
with full width $w=2$. Panels (a-c) show the magnitude of the order parameter $R$
as a function of $\beta$ for constant $\alpha$ scans from
numerical simulations of (\ref{AmplitudePhase}) for 1000 oscillators and $K=0$.
Limits of the unsynchronized state are consistent with
the linear stability analysis.
For the largest value shown, $\alpha=0.9$, the low $\beta$
transition $\beta=\beta_{c1}\simeq0.8$ is weakly hysteretic,
whereas the large $\beta$ transition $\beta<\beta_{c2}=3.7$ is continuous.
The state growing for $\beta<\beta_{c2}$ is a novel state with
$R\neq0$, but with no oscillator frequency locked to the order parameter, which
has a frequency outside of the band of shifted oscillator frequencies.
For $\beta>1.8$ there is also a state with $R$ close to unity in which all or most
of the oscillators are locked to the order parameter. For smaller
$\alpha=0.42$ there is a stable
small $R$ state for $\beta_{c1}<\beta<\beta_{c2}$, as well as a large $R$
solution. For $\alpha=0$ the large $R$ synchronized state persists down to
$\beta>1.6$, whilst the unsynchronized state remains linearly stable for all
$\beta$ (panel c). Panel (d) shows the phase diagram, including results from the
simulations as well as the linear stability analysis of the unsynchronized and fully
locked state. Over a large portion
of the $\alpha,\beta$ plane there are two stable solutions---a large $R$
synchronized state, and either the unsynchronized state (hatched region) or a
small $R$ state (cross hatched region)---leading to hysteresis for continuous
parameter scans. Over the dotted portion only a synchronized state is stable,
and over the unshaded region only the unsynchronized state is stable.

This material is based upon work supported by the National Science Foundation under
Grant No.~DMR-0314069, the U.S.-Israel Binational Science Foundation
Grant No.~1999458, the PHYSBIO program with funds from the European Union and NATO,
and HRL Laboratories, LLC.


\begin{thebibliography}{99}                                                                                               %

\bibitem {R01}M.~L.~Roukes, Scientific American \textbf{285}, 42 (2001).

\bibitem {C03}For a recent textbook, see A.~N.~Cleland, ``Foundations of
Nanomechanics,'' (Springer, Berlin, 2003).

\bibitem {HZMR03}X.~M.~H.~Huang, C.~A.~Zorman, M.~Mehregany, and M.~L.~Roukes,
Nature \textbf{421}, 496 (2003).

\bibitem {BR02}E.~Buks and M.~L.~Roukes, J.~Microelectromech.~Sys.~\textbf{11}%
, 802 (2002)

\bibitem {SHSICC03}M.~Sato, B.~E.~Hubbard, A.~J.~Sievers, B.~Ilic,
D.~A.~Czaplewski, and H.~G.~Craighead, Phys.~Rev.~Lett.~\textbf{90}, 044102 (2003).

\bibitem {LC03}R.~Lifshitz and M.~C.~Cross, Phys.~Rev.~B \textbf{67}, 134302 (2003).

\bibitem {PRK01}A.~Pikovsky, M.~Rosenblum, and J.~Kurths, ``Synchronization: A
universal concept in nonlinear science,'' (Cambridge University Press,
Cambridge, 2001).

\bibitem {SYNC}S.~H.~Strogatz, ``SYNC: The emerging science of spontaneous
order,'' (Hyperion, New York, 2003).

\bibitem {BSRW02}M.~Bennett, M.~F.~Schatz, H.~Rockwood, and K.~Wiesenfeld,
Proc.~Roy.~Soc.~Series A \textbf{458}, (2002)

\bibitem {W67}A.~T.~Winfree, J.~Theor.~Bio.~\textbf{16}, 15 (1967)

\bibitem {K75}Y.~Kuramoto,
Lecture Notes in Physics \textbf{39}, 420 (1975)

\bibitem {S88}H.~Sakaguchi, Prog.~Theor.~Phys.~\textbf{79}, 39 (1988)

\bibitem {MMS91}P.~C.~Matthews, R.~E.~Mirollo, and S.~H.~Strogatz, Physica D
\textbf{52}, 293 (1991)

\bibitem {AEK90}D.~G.~Aronson, G.~B.~Ermentrout, and N.~Kopell, Physica D
\textbf{41}, 403 (1990)

\bibitem {C00}H.\ G.\ Craighead, Science \textbf{290}, 1532 (2000)

\bibitem {BR01}E.~Buks and M.~L.~Roukes, Europhys.~Lett.~\textbf{54}, 220 (2001)
\end{thebibliography}
\end{document}